\documentclass[
twocolumn,
amsmath,
amssymb,
aps,
prl,
]{revtex4-2}

\usepackage{graphicx}
\usepackage{dcolumn}
\usepackage{bm}
\usepackage{braket}
\usepackage[dvipsnames]{xcolor}
\usepackage{color}

\newcommand{\magenta}{\textcolor{magenta}}
\usepackage{hyperref}
\usepackage{soul} 

\def\3he{$^3$He}
\def\4he{$^4$He}

\def\Aph{\textit{A} phase}
\def\Bph{\textit{B} phase}
\def\etal{\textit{et al.}}

\definecolor{tabgreen}{RGB}{44, 160, 44}
\definecolor{tabpurple}{RGB}{148, 103, 189}
\definecolor{taborange}{RGB}{255, 127, 14}

\begin{document}

\preprint{ASA16-SUS}

\title{Magnetic Susceptibility of Andreev Bound States in Superfluid \3he-B}

\author{J. W. Scott}
\email{johnscott2025@u.northwestern.edu}
\author{M. D. Nguyen}%
\author{D. Park}
\author{W. P. Halperin}
\email{w-halperin@northwestern.edu}
\affiliation{Northwestern University, Evanston, IL 60208, USA}

\date{\today}

\begin{abstract}
Nuclear magnetic resonance measurements of the magnetic susceptibility of superfluid \3he imbibed in anisotropic aerogel reveal anomalous behavior at low temperatures. Although the frequency shift clearly identifies a low-temperature phase as the \Bph, the magnetic susceptibility does not display the expected decrease associated with the formation of the opposite-spin Cooper pairs. This susceptibility anomaly appears to be the predicted high-field behavior corresponding to the Ising-like magnetic character of surface Andreev bound states within the planar aerogel structures.
\end{abstract}

\maketitle

Unconventional superconductors break symmetries beyond gauge symmetry, and are classified by the symmetries of the order parameter. The paradigm of unconventional superconductors is \3he{}, a spin-triplet $p$-wave BCS superfluid~\cite{Leg.73, Whe.75} with two zero field phases $A$ and $B$,  breaking and preserving time-reversal symmetry respectively. One identifying  characteristic of unconventionality is the strong suppression of the transition temperature, and concomitantly, the amplitude of the order parameter induced by nonmagnetic impurities~\cite{Lar.65}.  This is in stark contrast to conventional superconductivity~\cite{And.59}. In the case of $^3$He, dilute non-magnetic impurities of silica aerogel particles that are much smaller than the coherence length, reduce both the amplitude of the order parameter and the transition temperature~\cite{Por.95, Spr.95}, just as expected for an unconventional superconductor~\cite{Thu.96,Thu.98}.  The $^3$He quasiparticle scattering from these impurities produces surface bound states, Andreev bound states,~\cite{Sha.01} irrespective of whether  the surface scattering of  $^3$He quasiparticles is specular or diffuse.  Specularity can be achieved  by coating the surfaces with at least  two atomic layers of $^4$He~\cite{Hei.21}.  Under those conditions the bound states in $^3$He-$B$ are expected to be Majorana fermions~\cite{Miz.16}.  

The evidence for the existence of these Andreev bound states, both theoretical and experimental, and their relation to the topological character of the \Bph{}, have been reviewed by Mizushima \etal\ 2016~\cite{Miz.16}. Acoustic impedance measurements have resolved a density of states consistent with the existence of a Majorana cone as specularity is increased~\cite{Mur.09}. The bound states are also responsible for dissipation from vibrating wire devices in the low-temperature limit of the \Bph{}~\cite{Aut.20}. Theoretical interpretation of ion mobility measurements at the specular free surface are also consistent with their Majorana character~\cite{Tsu.17,Ike.19}. Particularly relevant for the present work are theoretical predictions for the  enhancement of the magnetic susceptibility~\cite{Chu.09,Nag.09, Nag.18} of the bound states.  Here, we report the detection of an anomalous enhancement to the magnetic susceptibility in the \Bph{} within a $98\%$ porosity silica aerogel with specular boundary conditions which we attribute to the magnetic susceptibility of \3he{} Andreev bound states.

\begin{figure*}
\includegraphics{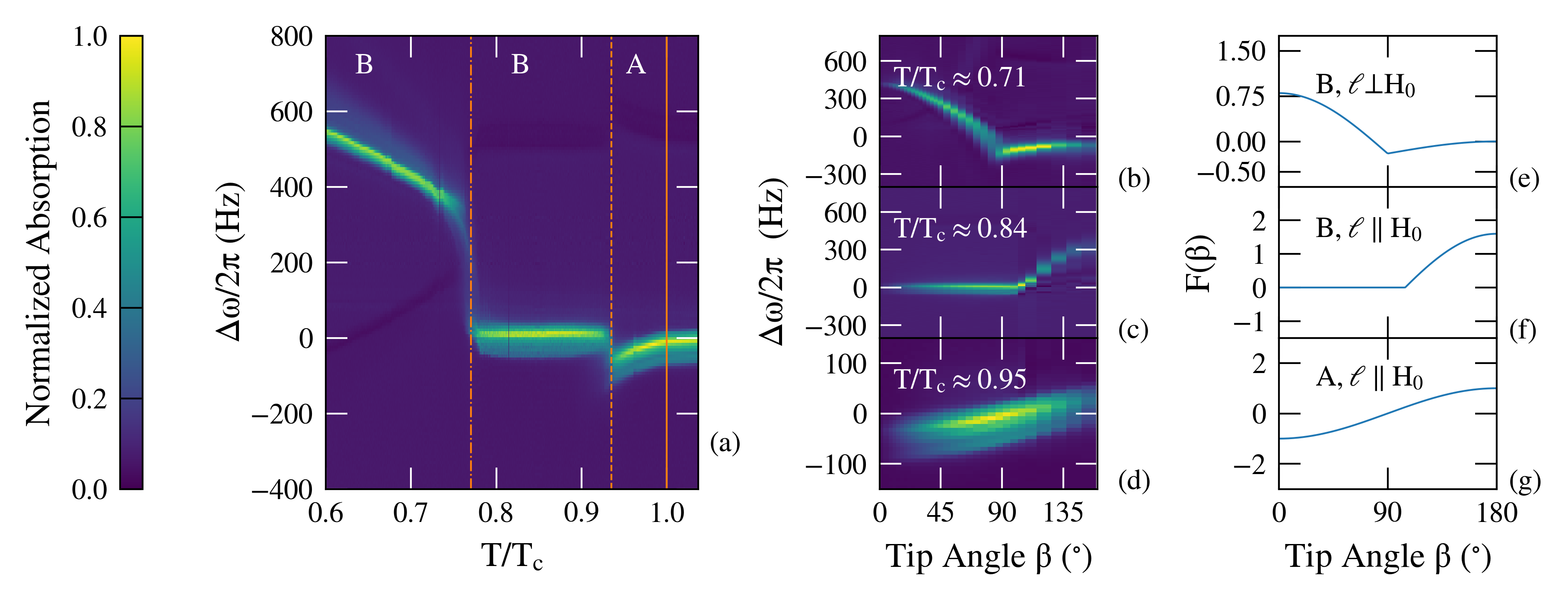}
\caption{\label{fig:3panewarmup} Temperature-dependence and tip angle dependence of NMR absorption spectra of \3he imbibed in $16\%$ stretched aerogel compared with pure \3he. In panels (a-d) the absorption spectrum is shown at each temperature and tip angle as a function of frequency by a color scale representing the spectral amplitude normalized to the maximum absorption amplitude in each panel. (a)\, Temperature dependence of spectra at tip angle, $\beta  = 8^{\circ}$, $P = 26.6$ bar, $H_{0} = 74.5$ mT. The configurations of the order parameter in order on warming; below the orbital flop transition $ T_{x} = 0.77 \, T_{c}$, $\bm{\ell} \perp \bm{H_{0}}$~\cite{Zim.18};  above $T_{x} = 0.77 \, T_{c}$, $\bm{\ell} \parallel \bm{H_{0}}$; above $T_{ab} = 0.93\,T_{c}$ the tip angle dependence is characteristic of the \Aph\ with $\bm{\ell} \parallel \bm{H_{0}}$. (b-d) Tip angle dependence of spectra in (a). (e-g) Theoretical tip angle dependence of spectra for pure \3he corresponding to (b-d)~\cite{Zim.19}.}
\end{figure*}

The application of aerogels for  investigation of $^3$He superfluid phases is now widespread~\cite{Hal.19}. It is significant that silica aerogels with global anisotropy~\cite{Ngu.23.arx}, couple directly to the $^3$He orbital angular momentum, producing well-defined, uniform order parameter textures~\cite{Li.14a,Zim.18}. The aerogel structure also determines the stability of different superfluid phases. Silica aerogel samples can be either grown, or mechanically strained, to produce uniaxial anisotropy~\cite{Pol.08}, stabilizing the \Aph{} when stretched (positive strain)~\cite{Pol.12} and the \Bph{} under compression (negative strain)~\cite{Li.14b}. In both phases  there is a second transition on cooling at $T_x < T_c$ from a uniform texture with angular momentum $\bm{\ell}$ parallel to the anisotropy axis, to a uniform texture with $\bm{\ell}$ perpendicular to the anisotropy axis\,~\cite{Ngu.23.arx}. We refer to this as the orbital flop transition\,~\cite{Zim.18}, clearly evident in Fig.\, \ref{fig:3panewarmup}(a) at $0.77 \,T_c$, where $T_c$ is the superfluid transition temperature. A different class of nematic alumina aerogels, Nafen, are extremely anisotropic; they stabilize the polar phase~\cite{Dmi.15a} which is a new $p$-wave superfluid phase that hosts half-quantum vortices~\cite{Aut.16}. In the present work we find that positive strain and specular boundary conditions for $^3$He quasiparticle scattering are requirements for the enhanced magnetic susceptibility that we observe in the superfluid $B$ phase.

We determine the magnetic susceptibility $\chi$ relative to the normal Fermi liquid magnetic susceptibility $\chi_{N}$ from the nuclear magnetic resonance (NMR) spectra of superfluid \3he\ imbibed in an anisotropic $16\%$ stretched aerogel (see the Appendix for details). Computational studies with a biased diffusion-limited cluster aggregation algorithm show that the microscopic structure of stretched aerogel has an anisotropic mean free path and planar voids~\cite{Ngu.23.arx}. Unlike in our prior experiments on similarly stretched aerogel, we have added $\approx 3.5$ layers of \4he{} to the surface, changing the quasiparticle scattering potential~\cite{Min.18,Hei.21} and eliminating the contribution of surface solid \3he{} to the spectral weight of the NMR signal~\cite{Col.09}. The addition of surface \4he{} has also been previously shown to alter the stability of superfluid phases imbibed in aerogel~\cite{Spr.96,Dmi.18,Zim.20}.


We identify a \Bph{} from the NMR frequency shift, Fig. \ref{fig:3panewarmup}. In the superfluid, macroscopic coherence of  the nuclear dipole-dipole interaction causes a  frequency shift $\Delta\omega$ away from the Larmor frequency $\omega_{L} = \gamma H_{0}$~\cite{Leg.73}. The frequency shift $\Delta\omega$ is given by,
\begin{equation}
\label{eq:freqshiftscale}
2 \omega_{L} \Delta \omega = \Omega(P, T)^{2} F(\beta).
\end{equation}
The magnitude of the frequency shift is set by the square of the longitudinal resonance frequency $\Omega^{2} \propto \Delta(P, T)^{2} / \chi$, where $\Delta(P, T)$ is the pressure- and temperature-dependent order parameter amplitude. The frequency shift dependence $F(\beta)$ on the NMR tip angle $\beta$ is set by the structure and orientation of the spin and orbital degrees of freedom of the order parameter~\cite{Zim.19}; further discussion of the NMR experiment, including the tip angle, is given in the Appendix. The theoretical behavior of $F(\beta)$ for pure \3he{} is displayed in Fig. \ref{fig:3panewarmup}\,(e-g), and can be compared with our data in panels (b-d) in this figure. The orbital degrees of freedom are defined by the  orbital angular momentum axis, $\bm{\ell}$, which is oriented by the aerogel anisotropy axis. In the present work the  static magnetic field $\bm{H_{0}}$ is aligned with the axis of anisotropy of the aerogel.

\magenta{}

At temperatures between $T_{c}$ and $T_{ab}$, as seen in Fig.\,\ref{fig:3panewarmup} (a), the nuclear magnetic resonance frequency decreases with decreasing temperature. This negative frequency shift is characteristic of the \Aph{} with $\bm{\ell} \parallel \bm{H_{0}}$. Its tip angle dependence is shown in Fig.\,\ref{fig:3panewarmup} (d), consistent with theory Fig.\,\ref{fig:3panewarmup} (g)~\cite{Zim.19}. This configuration results from anisotropic quasiparticle scattering induced by the planar aerogel structure~\cite{Ngu.23.arx} in a manner analogous to confinement in a slab~\cite{Aho.75a} and consistent with results from other anisotropic aerogels~\cite{Li.14a,Dmi.20a}. For comparison, in the pure bulk superfluid the energetically favored configuration is $\bm{\ell} \perp \bm{H_{0}}$ for which the frequency shifts are positive.

\begin{figure}[]
\includegraphics{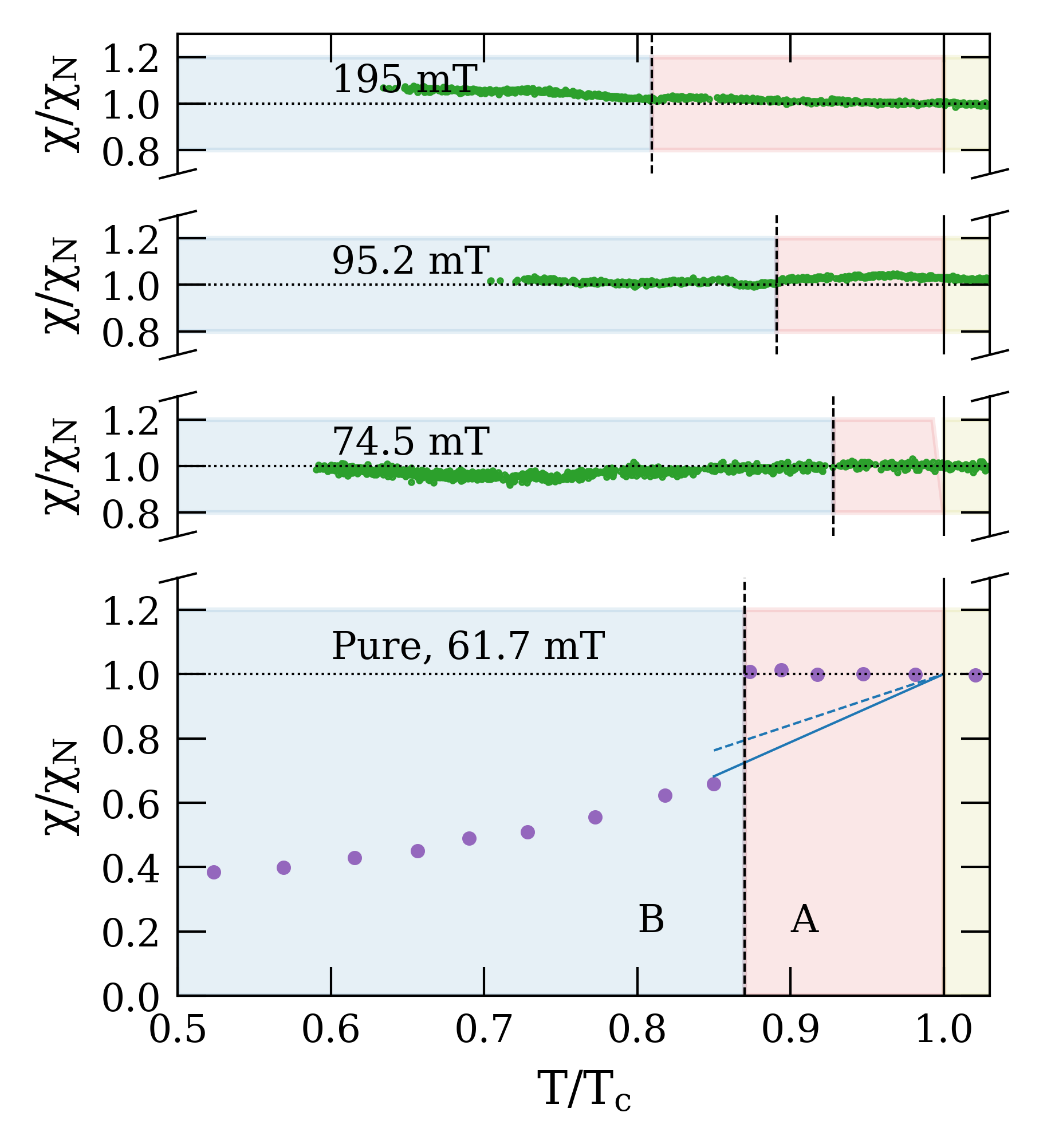}
\caption{\label{fig:susceptibility} Magnetic susceptibility $\chi / \chi_{N}$ (green circle), over a range of fields at P = 26.6 bar. In contrast to the behavior expected, the magnetic susceptibility is temperature independent in the \Bph{} (blue background) for all fields, different pressures, temperatures below $T_{ab}$, and different orbital configurations. For comparison, $\chi / \chi_{N}$ of the pure superfluid at $P = 27.0 $ bar is shown in the lowest panel (purple circle), taken from Ref.\,~\cite{Sch.81}. The blue solid (dashed) lines are the Born (unitary), Ginzburg-Landau results from the homogeneous isotropic scattering model for \3he{} in aerogel~\cite{Sha.01}.}
\end{figure}

At lower temperatures, the superfluid enters the \Bph{} which we identify from its characteristic NMR frequency shifts, shown in Fig. \ref{fig:3panewarmup} (a). The shifts above and below $T_{x} = 0.77\,T_c$ are characteristic of the \Bph{} with $\bm{\ell} \parallel \bm{H_{0}}$, Fig.\,\ref{fig:3panewarmup} (c,f) and $\bm{\ell} \bot \bm{H_{0}}$, Fig.\,\ref{fig:3panewarmup} (b,e) respectively, where  $T_{x}$ is the orbital flop transition~\cite{Zim.18}, referred to earlier. The magnetic susceptibility in  the \Bph{} is temperature independent within our measured temperature range, Fig.\,\ref{fig:susceptibility}, in striking contrast to the temperature dependent magnetic susceptibility in the pure superfluid.

In pure \3he{} $B$, the reduction of the magnetic susceptibility with decreasing temperature corresponds to the formation of opposite-spin Cooper pairs~\cite{Bal.63, Leg.65a}. This is observed in both pure and impure superfluids~\cite{Sch.81, Spr.96}. The extent of the reduction is less in a large magnetic field~\cite{Fis.86, Ran.96}, or in confinement,~\cite{Lev.13a} which tends to suppress the formation of opposite-spin pairs. The effect corresponds to a polar distortion of the \Bph{} order parameter that can be measured independently.  We follow the method of Rand {\it et al.}~\cite{Ran.96}, finding that the distortion varies smoothly from $\sim 0.34$ to $0.09$ in the region below $ 0.77 \, T_{c}$, as shown in Fig.\,\ref{fig:distortion} (see Appendix). The  polar distortion  $\leq 0.09$ increases the susceptibility in the pure \Bph{} by $\leq 0.15 \, \chi / \chi_{N}$~\cite{Fis.86,Ran.96}.  Polar distortion cannot account for the excess susceptibility. Similarly, the effect of  quasiparticle scattering from impurities is also too small as shown in Fig.\ref{fig:susceptibility}.


\begin{figure}[]
\includegraphics{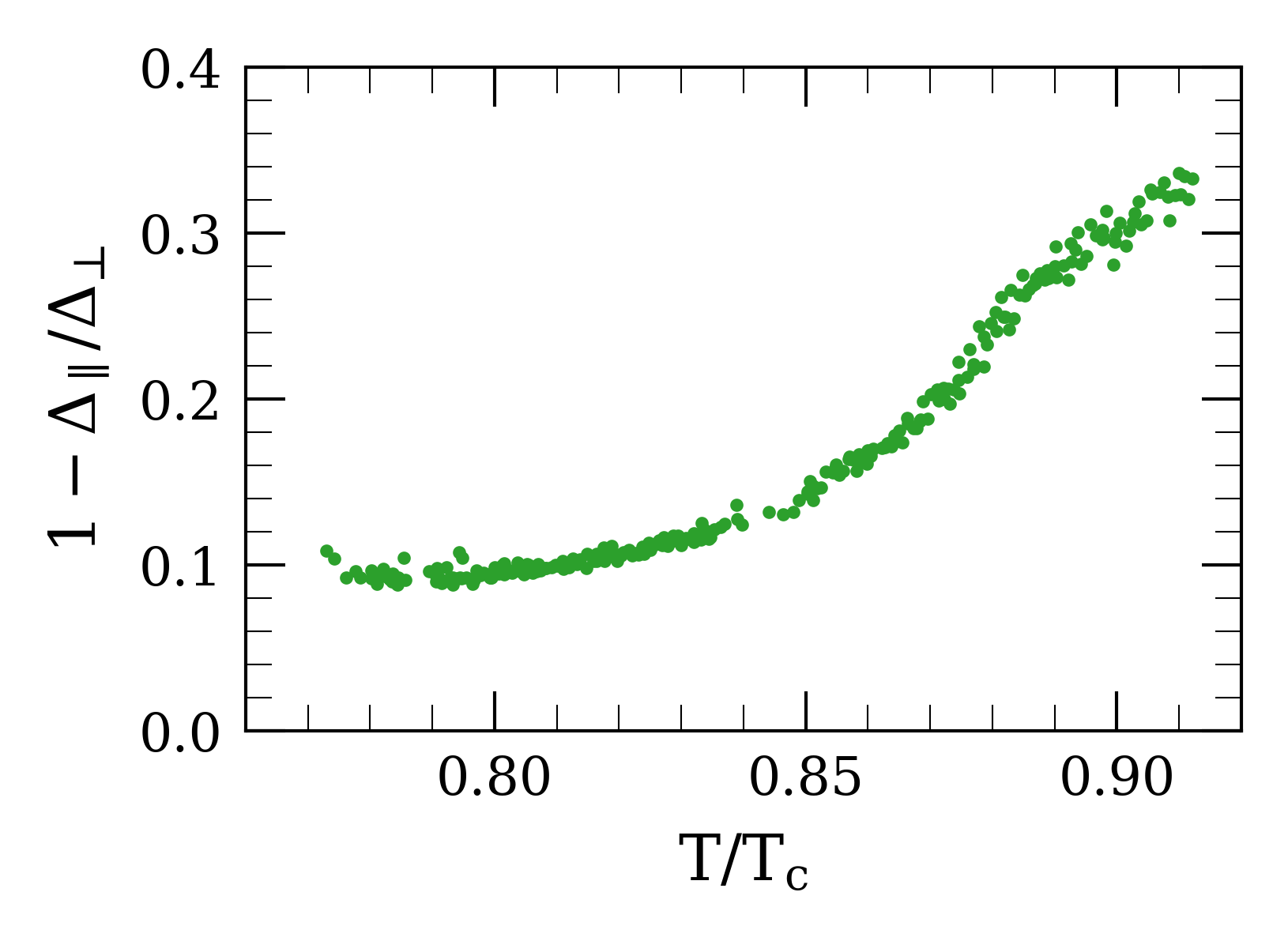}
\caption{\label{fig:distortion} Measured order parameter distortion. The order parameter amplitude distortion in the direction of the magnetic field, $1-\Delta_{\parallel} / \Delta_{\bot}$ at $H_{0} = 74.5$ mT for the temperature range between $T_{ab}$ and $T_{x}\, = 0.77 \, T_{c}$. The degree of order parameter distortion $\approx 0.34$ close to $T_{ab}$ is comparable to that seen in nanoconfined planar slabs~\cite{Lev.13a}. At lower temperatures the distortion we measure $\approx 0.09$ approaches the value for  bulk, pure \3he{} $\approx 0.005$ in a $112$ mT magnetic field~\cite{Ran.96}.}
\end{figure}

The earliest magnetic susceptibility experiments on the impure \Bph{} in a nominally isotropic aerogel, were performed by Sprague \etal{}~\cite{Spr.96} with \4he{} covering the surface of the aerogel, similar to the experiments we report here. They found a small increase in the susceptibility compared to the pure superfluid state, subsequently accounted for by microscopic theories for isotropic impurities including polarization of Andreev bound states~\cite{Sha.01, Min.01}. In the treatment by Sharma and Sauls~\cite{Sha.01}, the aerogel was modeled as uncorrelated impurities with a single mean-free path, from which both susceptibility and suppression of the critical temperature were determined. This theory has also been used to account for the temperature dependent susceptibility observed in both isotropic~\cite{Pol.11} and compressed aerogel samples~\cite{Li.14b} without \4he{} preplating, which do not exhibit the anomalous behavior we report here. The susceptibility in the Ginzburg-Landau limit of the quasiclassical theory is shown in Fig.\,\ref{fig:susceptibility} as a blue solid (dashed) line for the Born (unitary) scattering limits. 

Theoretical studies of the magnetic susceptibility from bound states in the \Bph{} near a solid plane boundary have found two effects, the emergence of a spontaneous polarization normal to the surface of the plane~\cite{Chu.09} and the enhancement of the magnetic susceptibility in a confined slab~\cite{Nag.09}. In the latter case, it was predicted that confining the \Bph{} to a slab causes an increase in the susceptibility, with the highest levels of confinement recovering the susceptibility of the normal liquid~\cite{Nag.09, Miz.12b}. A detailed study of the field dependence of Andreev bound states found that the susceptibility increase is associated with a gap in the Andreev bound state dispersion spectrum induced by magnetic field~\cite{Miz.12a}.

The NMR frequency shift of the superfluid below $T_{ab}$ precludes the identification of the superfluid state as any of the experimentally observed or theoretically well-established equal spin pairing states, such as the polar or planar phases. The magnetic susceptibility also does not display the decrease characteristic of the isotropic impure superfluid models, and is inconsistent with the Ginzburg-Landau limit of these theories. Simulations of the growth of anisotropic aerogels~\cite{Ngu.23.arx}, together with small angle x-ray scattering~\cite{Pol.08} provide strong evidence for the existence of planar structures in stretched aerogels. As a consequence, we propose that planar structures in the aerogel are responsible for the enhanced susceptibility in a manner analogous to the predictions for Andreev bound states in confined slabs.


\begin{figure}[b]
\includegraphics{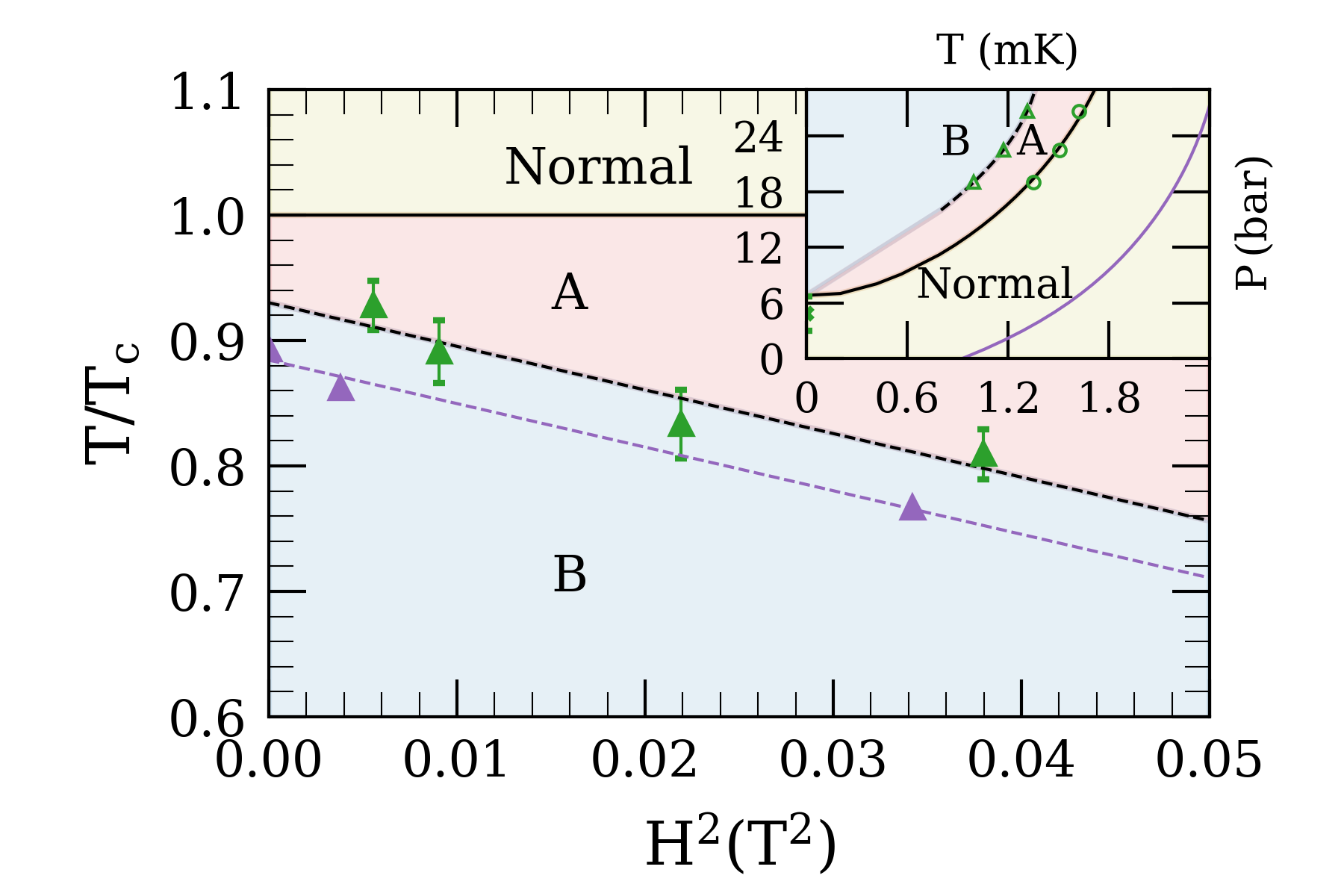}
\caption{\label{fig:phase_diagram} Magnetic phase diagram at $P = 26.6$ bar. The green trace (green triangle) denotes the experimentally observed $T_{ab}$ in the aerogel. A linear fit shows the extrapolated $T_{ab}$ at zero field to be $T_{ab} = 0.93\,T_{c}$. The purple trace (purple triangle) denotes the $A-B$ transition in pure bulk superfluid at $P=27.0$ bar~\cite{Gre.86, Sch.81} for comparison. (Inset) The pressure-temperature phase diagram at  $H_{0} = 195$ mT, with  critical temperature $T_{c}$ (green open circle), A-B transition temperature $T_{ab}$ (green open triangle). The black curve is a fit for the observed $T_{c}$ based on an impurity model~\cite{Thu.96, Sau.03, Aoy.06}. The purple line denotes the pure liquid $T_{c}$, the dashed black line is a guide to the eye.}
\end{figure}

According to theory, a slab of thickness of $\approx 5\,\xi_{0}$ is required to recover the full normal state susceptibility from bound states~\cite{Nag.09}. In the pressure range of the current work this corresponds to $130$ nm taking the zero temperature cohrence length to be $\xi_{0} =(\hbar v_{F})/(2 \pi k_{B} T_{c})$;  $v_{F}$ is the Fermi velocity. Numerical simulations of the stretched aerogel structure  indicate an approximately  planar mass distribution with plane separations of $\sim 50-100$\,nm.  This is  similar to the calculated quasiparticle mean free path of $90-120$\,nm~\cite{Ngu.23.arx}.  Although aerogel and slab confinements are very different, the slab thickness for the expected enhanced magnetic susceptibility is roughly the same as that of the planar structure of the stretched aerogel.

The $A$ to $B$ transition temperature, $T_{ab}$, decreases with a quadratically increasing field with a slope comparable to that of the pure superfluid and similar for all isotropic aerogels and anisotropic silica aerogels with $^4$He preplating including the present work, and shown in Fig. \ref{fig:phase_diagram}~\cite{Tan.91,Sch.81,Ger.02,Bhu.10,Li.15, Zim.20}. This slope taken at comparable pressures in aerogel samples with and without preplating is included in Supplementary Material as Table 1\cite{Pol.11, Li.14b, Zim.20}. Since this transition is first-order its field dependence is given by a Clausius-Clapeyron relation between the differences in susceptibility $\chi$ and entropy $S$ for the two superfluid phases,
\begin{equation}
\label{eq:clausiusclapeyron}
\frac{d T_{ab}}{d H^{2}} = -\frac{1}{2} \frac{\chi_{A} - \chi_{B}}{S_{A} - S_{B}}.
\end{equation}
Our measurements of the magnetic susceptibility $\chi_B$ are in stark contrast with the Clausius-Clapeyron relation. 

In summary, we have discovered an anomalous contribution to the susceptibility of superfluid \3he{}-B imbibed in a stretched aerogel. This susceptibility appears to be a consequence of planar regions in the aerogel structure and is surprisingly large, essentially identical to that of the normal state, similar to predictions for the susceptibility of \3he{}-B confined to a planar slab with specular boundary conditions \cite{Nag.09}. How this can be reconciled with the field dependence of the $AB$ transition remains an open question.

This work was supported by NSF Division of Materials Research grant DMR-2210112. We are grateful to J. A. Sauls and A. M. Zimmerman for useful discussion and to V.P. Mineev for communication. We also thank the Northwestern University Instrument Shop for use of its facilities.

\section{Appendix on Experimental Methods}

We grow the $16$\% stretched aerogel with the rapid supercritical extraction method presented in Pollanen \etal {}~\cite{Pol.08}. The aerogel is grown in a $4.20$ mm inner diameter glass NMR tube, however, during the growth process the aerogel shrinks radially inwards from the walls of the NMR tube. The NMR sample is approximately $3.53$ mm in diameter and it measures $5.5$ mm long. Density measurements show the sample to be approximately $97.5 \%$ porous. Optical birefringence measurements reveal an axis of structural anisotropy aligned with the cylindrical axis of the aerogel. The cylindrical axis is oriented along the direction of the static NMR field $H_{0}$, and connected by a fill line to the silver heat exchanger of the cryostat.

The NMR experiments require a radio frequency pulse of length $\tau_{p} \propto \beta$, the proportion calibrated in the normal Fermi liquid by observing the dependence of the intensity of the NMR signal on the length of the pulse $\tau_{p}$; maximal intensity corresponds to $\beta = 90^{\circ}$. After cooling by nuclear demagnetization, intermittent NMR pulses of several fixed lengths are delivered as the sample warms, producing the color map in Fig.\,\ref{fig:3panewarmup} (a). During this process we measure temperature via a capacitive pressure transducer \3he{} melting curve thermometer~\cite{Gre.86} and a $^{195}$Pt susceptibility thermometer. The cryostat can also be held at a roughly fixed temperature while spectra are captured for a variety of pulse lengths, producing the color maps in Fig.\,\ref{fig:3panewarmup} (b-d). Each  NMR spectrum in the temperature sweep data is accumulated over a small region of temperature, typically $\sim 8$\,$\mu$K.

Measurements were performed between pressures of 19.0 bar and 26.6 bar and magnetic fields from 74.5 to 195 mT. The pressure-dependent critical temperature suppression of the $16\%$ stretched aerogel is consistent with a quasiparticle mean free path $\lambda = 131$ nm and aerogel correlation length $\xi_{a} = 23$ nm~\cite{Sau.03}.

We determined the polar distortion of the amplitude of the order parameter, which is the suppression of $\Delta_{\parallel}$ aligned with the magnetic field relative to the perpendicular component $\Delta_{\bot}$, following the method of Rand \etal~\cite{Ran.96}. The frequency shift was measured at two tip angles above and below the critical tipping angle (nominally $\arccos(-1/4) \approx 104^{\circ}$), separating the regions where $\Delta \omega$ is small and large (e.g. $10^{\circ}$ vs. $\approx 135^{\circ}$ in Fig. \ref{fig:3panewarmup} (c)). The frequency shifts for two different pulse lengths taken on warming determine the distortion of the order parameter as a function of temperature. This gap distortion is given by,

\begin{equation}
\label{eq:gap_disto}
\frac{\Delta_{\parallel}}{\Delta_{\bot}} = \frac{1 + \frac{\Delta \omega_{1}}{\Delta \omega_{2}} + 2 \frac{\Delta \omega_{1}}{\Delta \omega_{2}} \cos{\beta_{2}}}{1 - 2 \frac{\Delta \omega_{1}}{\Delta \omega_{2}} \cos{\beta_{2}}}
\end{equation}
with $\Delta \omega_{1}$ and $\Delta \omega_{2}$ the frequency shifts of the small and large pulses, respectively, and $\beta_{2}$ the tipping angle of the large pulse\cite{Has.83}. This ratio is independent of the magnetic susceptibility of the superfluid and the \Bph{} longitudinal resonance frequency $\Omega_{B}$.

The susceptibility is proportional to the integral of the absorption spectrum of the NMR signal. At temperatures above those shown in Fig. \ref{fig:susceptibility} there is an additional contribution to the susceptibility from the \3he{} in the fill lines and from that in the near vicinity of the aerogel, which we refer to as the \3he{} bath. At temperatures substantially higher than $T_{c}$ in the aerogel, the bath is in the superfluid \Aph{}. As a consequence, the contribution of the bath to the signal, $\sim 20\,\%$ of the total spectral weight, is shifted in frequency away from the signal of the normal liquid in the aerogel, as shown in Supplemental Material Fig. 1. At and below $T_{ab}$ in the pure superfluid, the \Bph{} bath signal abruptly becomes very broad.  At lower temperatures, below $T_c$ in the aerogel as shown Fig. \ref{fig:susceptibility}, this contribution cannot be distinguished from the background noise. The influence of the superfluid bath on the susceptibility in both stretched and compressed aerogel can be seen in Supplemental Material Fig. 1\cite{Zim.18, Li.13}.

\bibliography{Scott_References}

\end{document}


\preprint{ASA16-SUS}

\title{Magnetic Susceptibility of Andreev Bound States in Superfluid \3he-B Supplementary Material}

\author{J. W. Scott}
\author{M. D. Nguyen}
\author{D. Park}
\author{W. P. Halperin}
\affiliation{Northwestern University, Evanston, IL 60208, USA}

\date{\today}

\maketitle

\begin{figure*}[ht!]
\includegraphics{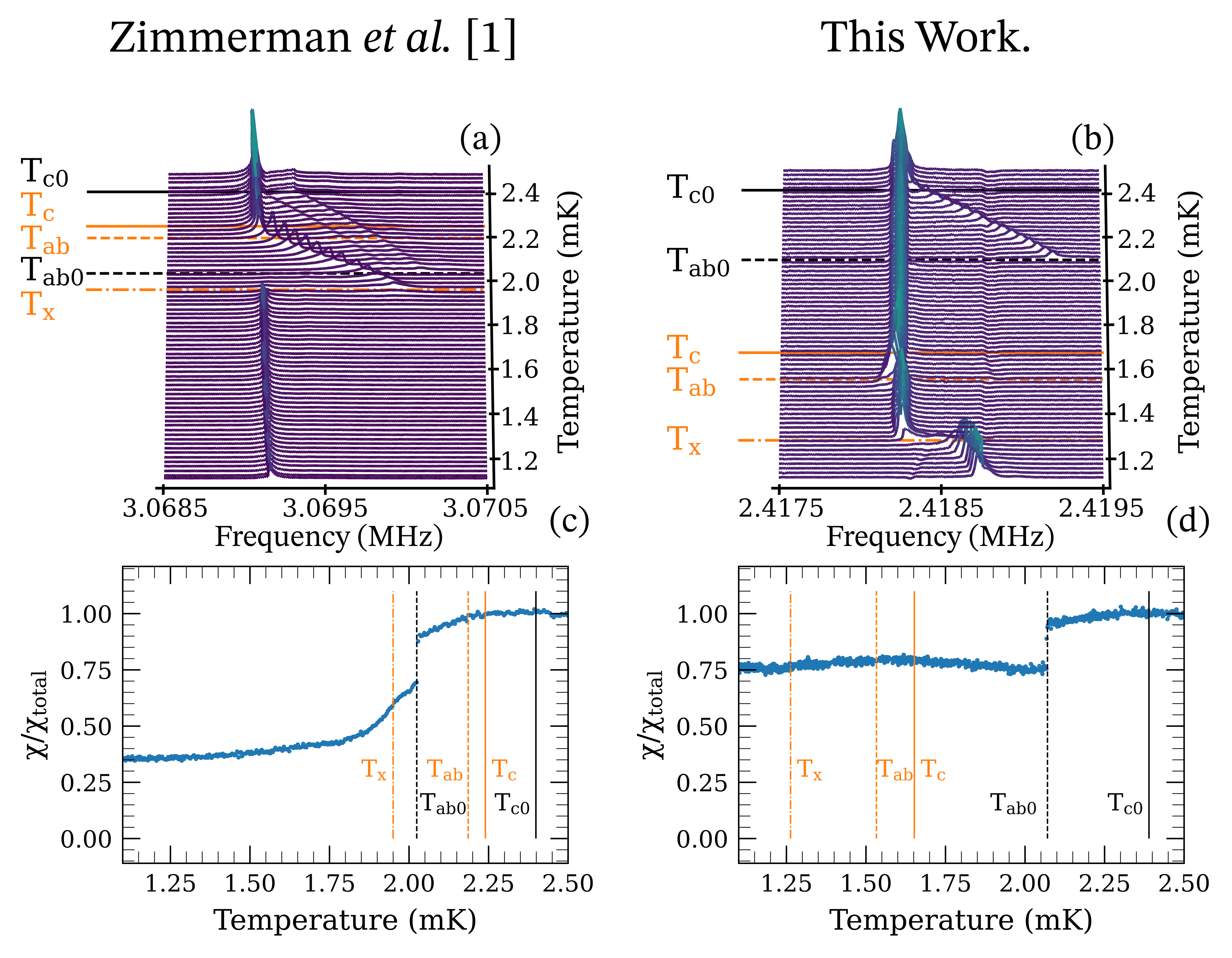}
\caption{\label{fig:4panel} Temperature dependence of the NMR absorption spectra at small tipping angle in (a) $19 \%$ compressed aerogel at 27 bar, $H_{0} = 0.096$ T with \4he{} preplating~\cite{Zim.18}; (b) $16\%$ stretched aerogel at 26.6 bar, $H_{0} = 0.0745$ T. The samples display frequency shifts characteristic of the order parameter orientations induced by  opposite strain. The order parameter orientations for the compressed aerogel on warming from within the  \Bph{}, are consecutively, $\bm{\ell} \parallel \bm{H_{0}}$, $\bm{\ell} \bot \bm{H_{0}}$, and  then a 2D-disordered \Aph{}~\cite{Li.13,Zim.18}. The configurations of the order parameter in (b) are explained in the main text. Figures (c) and (d) display the full temperature dependent susceptibility of the NMR absorption spectra shown in (a) and (b) respectively, including the superfluid bath. The solid vertical lines are the superfluid critical temperature in the  bath ($T_{c0}, \,\textrm{--}$) and in the aerogel, ($T_{c}, \, \textcolor{taborange}{\textrm{--}}$). Likewise, the dashed vertical lines are for the bath and for the aerogel $A$ to $B$ transitions, ($T_{ab0},\,\textrm{--}$),  ($T_{ab}, \,\textcolor{taborange}{\textrm{- -}}$) . The dash-dotted vertical lines are the orbital flop transitions ($T_{x},\,\textcolor{taborange}{\textrm{-}\cdot}$)~\cite{Zim.20}. As seen in the figure above, the combination of linewidth broadening from the much larger distribution of frequency shifts of the \Bph{} superfluid bath and its reduced susceptibility,  cause the spectrum to quickly disappear into the noise floor of the spectrometer below $T_{ab0}$. In panel (b) a small negative contribution to the NMR spectrum appears. This is an artifact of the spectrometer when operated at low frequencies and is mirrored across the NMR pulse carrier frequency $\omega_{c}/(2\pi) = 2.4185$ MHz (see Appendix). No attempt to subtract or correct this artifact has been made.}
\end{figure*} 

\clearpage

\begin{table}[hb!]
\begin{tabular}{|c|c|c|c|}
\hline Strain & \4he{} Preplating & $(dT_{ab}/T_{c})/(dH^{2})$ ($T^{-2}$)& P (bar)\\ 
\hline 
+$16 \%$ & Yes & $-3.473$ & 26.5 \\
0 & No & $-3.644$  & 26.3 \\
-$12 \%$ & No & $-3.725$ & 26.5  \\
-$19 \%$ & Yes & $-4.256$ &  27 \\
-$19 \%$ & No & $-4.741$ & 26.3  \\
-$20 \%$ & No & $-4.521$ & 26.5 \\
-$30 \%$ & No & $-5.161$ & 26.5  \\
\hline
\end{tabular}
\caption{\label{tab:g_beta} A record of slopes for the field dependence of the $A$ to $B$ transitions for different aerogel samples at comparable pressure, as presented in Fig. 4 in the main text\cite{Pol.11, Li.14b, Zim.20}. For aerogel samples with no or small strain, the slope of the field dependence of $T_{ab}(H^2)$ in the form of $\frac{d T_{ab}/T_{c}}{d H^{2}}$ approximates that of the bulk value, $-3.477$ $T^{-2}$, independent of preplating. In aerogel samples with a large negative strain and no preplating, the slope of the transition is substantially steeper. The steeper slope corresponds to the appearance of a critical transition field $H_{c}$ in these compressed samples~\cite{Li.14b}.}
\end{table}

\bibliography{Scott_References}